# PHENOMENOLOGICAL MODELING OF THE LIGHT CURVES OF ALGOL-TYPE ECLIPSING BINARY STARS

Ivan L. Andronov

*Department "High and Applied Mathematics", Odessa National Maritime University, Odessa, Ukraine, e-mail: tt_ari @ ukr.net*

*We introduce a special class of functions for mathematical modeling of periodic signals of special shape with irregularly spaced arguments. This method was developed for determination of phenomenological characteristics of the light curves, which are necessary for registration in the "General Catalogue of Variable Stars" and other databases. For eclipsing binary stars with smooth light curves – of types EB and EW – it is recommended a trigonometric polynomial of optimal degree in a complete or symmetric form. For eclipsing binary systems with relatively narrow minima (EA-type), statistically optimal is an approximation of the light curves in a class of non-polynomial spline functions. It is used a combination of the second-order trigonometric polynomial (TP2, what describes effects of "reflection", "ellipsoidality" and "spotness") and localized contibutions of minima (parametrized in depth and profile separately for primary and secondary minima). Such an approach is characterized by a statistical accuracy of the smoothing curve, which is up to ~1.5-2 times better than the trigonometric polynomial of statistically optimal degree, and the absence of false "waves" in the light curve associated with the effect of Gibbs. In addition to the minimum width, which can not be determined by a trigonometric polynomial approximation, the method allows to determine with better accuracy its depth, as well as to separate the effects of the eclipse and out-of-eclipse parts. For multi-color observations, improving the accuracy of the smoothing of the curve in each filter will allow to obtain with better accuracy the curves of the color index variations. Effectivity of the proposed method increases with decreasing eclipse depth. The method is a complementary one to the approximation using trigonometric polynomials. The method called NAV ("New Algol Variable"), was illustrated by application to eclipsing binary systems VSX J022427.8-104034=USNO-B1.0 0793-0023471 and BM UMa. For VSX J022427.8-104034, an alternative model of «double period» is discussed.*

### Introduction

The study of variable stars is one of the most important areas of modern astrophysics. For the classification of new variable stars discovered and entering them in the "General Catalogue of Variable Stars" [1] (a regularly updated electronic version is available at *http://www.sai.msu.su/groups/cluster/gcvs/gcvs/*) or other databases (eg. "Variable Star Index", *http://www.aavso.org/vsx*, or national or observatory lists of newly discovered variable stars), one must determine the characteristics of the variability. In particular, to search for possible periodicity, using a large number of methods for periodogram analysis. They can be divided into groups of "nonparametric" (or "point-point") and "parametric" (or "point-line"). To study the non-parametric method of Lafler and Kinman [2], improved by Kholopov [3] and its modifications, the comparative analysis which was carried out by Andronov and Chinarova [4].

The parametric methods are based on the analysis of deviations of individual observations ("points") from an approximating curve ("line"), the parameters of which are determined using the least squares method (e.g. monograph by Anderson [5]) . The problems of extraction of periodic components were discussed in monographs by Terebizh [6,7].

In this paper, we introduce an algorithm of approximation of periodic signals using basic functions of special shape, specially for the Algol-type (EA) stars. These are «binaries with spherical or slightly ellipsoidal components. It is possible to specify, for their light curves, the moments of the beginning and end of the eclipses» [8,1,9], contrary to types EB and EW. From the mathematical point of view, this means that the variations in EB and EW are smooth (the derivative is continuous), but in EA the derivative drastically changes at beginning and end of the eclipse. Thus for the types EB and EW the coefficients of the trigonometric polynomial decrease fast with the harmonic number, and the statistically optimal degree *s* is relatively small [10,11].

With decreasing width of the minimum *D*, which is an optional parameter for this type of stars in the "General Catalogue of Variable Stars" [1], the value of *s* increases, what causes worsening of the approximation quality – either due to excitation of formal «waves» at the light curve (the Gibbs phenomenon [12]), or due to an increase of statistical error of approximation [10,11].

«Phenomenological» modeling is an optimal one for the majority of known binary stars, because the «physical» modeling (e.g. monograph [13]) based on the method of Wilson-Devinney [14] and its numerous modifications needs an additional information on temperatures and radial velocities of both components and on a mass ratio. Such an information is available only for ~1% of objects. For other ~99%, the « phenomenological » models remain to be important.

The determination of physical charactersistisc of components (and maybe spots) needs to solve an incorrect inverse problem (e.g. [15]) and typically causes an extended region of solutions in the parameter space, which produce similar results in a quality of approximation. As is shown in this work, the

«phenomenological» modeling agrees with «physical» approximation within accuracy estimate, but needs significantly smaller computational resources.

**Mathematical Models**

*Trigonometrical Polynomial*

For periodic signals, the realation $x(t+kP) = x(t)$ is valid, where $x(t)$ - the sognal value at time $t$, $P$ – period and $k$ – an arbitrary integer number. If the measured signal is brightness (stellar magnitude), the graph of dependence $x(t)$ of the observed signal values $x_i$ on time $t_i$ ($i=1..n$ – the number of a current observation) is called "the light curve". Because typically the observations of astronomical objects are carried out not continuously, but at discrete moments of time (e.g. patrol photographicall, visual or CCD monitoring), such a presentation is not suitable. For the stars with mono-periodic variations, it is possible to write a general expression using the trigonometric polynomial of degree $s$ (which is sometimes called the "restricted Fourier series" [13]):

$$x_{TP,s}(t) = C_1 + \sum_{j=1}^{s}(C_{2j}\cos(j\omega(t-T_0)) + C_{2j+1}\sin(j\omega(t-T_0))) = \qquad (1)$$

$$= C_1 + \sum_{j=1}^{s} R_j \cos(j\omega(t-T_{0j})) \qquad (2)$$

Here the coefficients $C_\alpha$, $\alpha=1..m$, are unknowns, $m=1+2s$, $\omega=2\pi/P$ – main circular frequency, $j\omega$ - the circular frequency of the harmonic number ($j$-1), $T_0$- «initial epoch» (this approximation is not dependent on it). Larger physical sense has a second form, where is used the semi-amplitude $R_j$ of the wave with circular frequency $j\omega$ and an «initial epoch» $T_{0j}$, which corresponds to a maximum value of the contribution of corresponding wave.

Is is noticeable that for symmetric light curves ($x(T_0+\tau)=x(T_0-\tau)$), it is effective to use an approximation with $m=1+s$ coefficients and $R_j=C_{j+1}$:

$$x_{TP,s}(t) = C_1 + \sum_{j=1}^{s} C_{j+1}\cos(j\omega(t-T_0)) \qquad (3)$$

However, the problem on a presence of asymmetry is usually solved by an estimate of statistical significance of the coefficients $C_{2j+1}$ at sine terms in Eq. (1), with an initial epoch $T_0$, corresponding to the minimum of the smoothing curve.

Taking into account that the maximum of stellar magnitude corresponds to minimum of flux, this sometimes leads to misunderstandings in the articles. For example, in the studies of pulsating stars, the initial epoch corresponds to the maximum brightness (ie, minimum stellar magnitude). Therefore, in [10,11] before the sum is the "minus" sign. However, for the eclipsing star's, the minimum brightness corresponds to the maximum stellar magnitude, so in this article, it is more convenient to use the definition (2).

For fixed value of the degree of the trigonometrical polynomial $s$ and in a general case of irregular arguments of the signal, the coefficients $C_\alpha$ are to be determined using the least squares method (e.g. [16]). In practice, in many articles are used simplified formulae for the «Digital Fourier Transform» (DFT), which produce values of the coefficients, which differ from statistically optimal ones, determined using the least squares.

The «symmetric» form (Eq. (3)) is used for investigation of dependencies of the amplitudes $R_j$ on physical characteristics of eclipsing binary systems (e.g. [17]).

For the illustration of the search of statistically optimal degree of the trigonometrical polynomial $s$ for the system of the EA-type, we used the observations of the eclipsing binary star VSX J022427.8-104034 = USNO-B1.0 0793-0023471 (hereafter VSX0224) in the Cetus constellation, which was discovered by Natalia A.Virnina.

The observations were published at the AAVSO site *http://www.aavso.org/vsx/docs/239151/1027/var_photometry.txt*. The discovery report was published in the electronic VSNET-circular *http://ooruri.kusastro.kyoto-u.ac.jp/mailarchive/vsnet-ecl/3749*, where, using our program [10] and the parameter $s=9$, were determined the values of the period $P=0.522734\pm0.000015^d$, initial epoch $T_0= 2455106.3231\pm0.0005^d$, smoothed values of brightness at minimum $m_{min}=16.079\pm0.008^m$ (no filter, the calibration in the photometric system R) and at minimum $m_{max}=15.417\pm0.011^m$.

In Fig 1. there are shown dependencies of the characteristics of approximating functions on the degree $s$ of the trigonometrical polynomial (see [10,11,16] for details). Using different criteria, the optimal values are $s=16$ (the Fischer's criterion with a false alarm probability FAP<0.01; minimal statistical accuracy of the determination of the photometric period) and $s=7$ (minimal statistical accuracy of determination of the smoothing function; of brightness at the minimum; of the phase of the minimum).

This result confirms that the Fischer's criterion often causes an overestimated value of the optimal degree $s$, what may be due to correlation between the deviations of observations, which were obtained at close times, on the smoothing curve.

The value $s=9$, which was used by N.A.Virnina, corresponds to significantly smaller value of FAP<$2\cdot10^{-8}$ (-lg FAP=7.7). However, the analysis of the light curves shows that, as the minimum at the light curve is narrow, the approximations either for $s=7$, or for $s=9$ are much higher than the observational points, and $s=16$ corresponds to much better description of the minimum itself, whereas there occur "waves" at the out-of-eclipse part of the light curve, due to the Gibbs phenomenon.

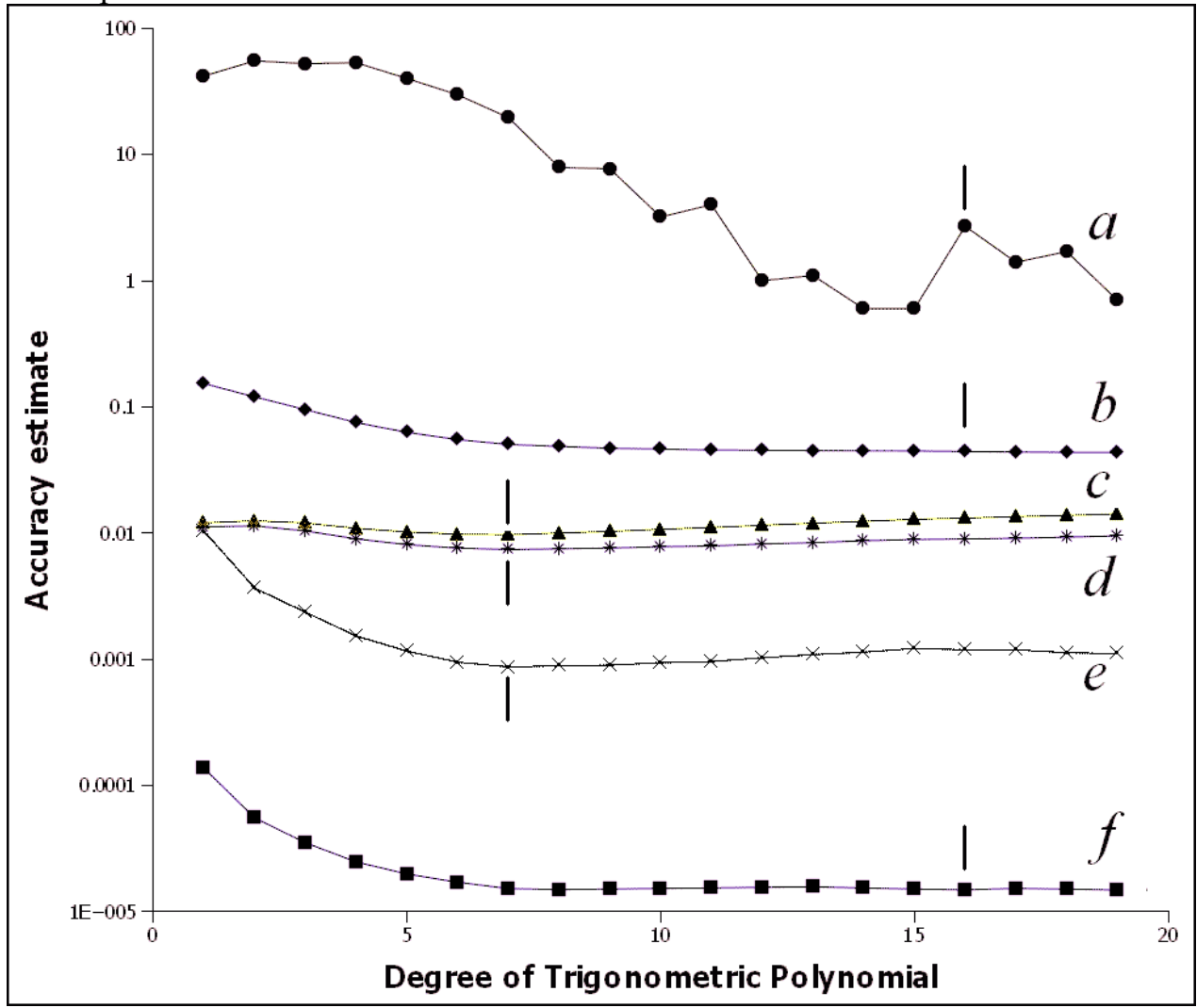

Fig.1. Dependence on the degree of the trigonometric polynomial $s$ of the characteristics of the approximation of the light curve of the eclipsing star VSX0224: a) –lg FAP, b) r.m.s. deviation of observations from the smoothing function; c) r.m.s. value of the accuracy of the smoothing light curve; d) accuracy of determination of the brightness at the minimum; f) accuracy of the period determination. Vertical bars correspond to the optimal (for a given criterion) value of $s$.

Thus hereafter, for the construction of the light curves of VSX0224, we use the values $T_0$= HJD 2455106.3236±0.0006, $P$=0.522724±0.000014$^d$, which correspond to $s=16$. The value of brightness at the minimum $m_{min}=16.109\pm0.009^m$ corresponds to the observations much better.

*Method of «Asymptotic Parabolae »*

For an independent estimate of parameters of the minimum at the phase curve, we have used the method of «asymptotic parabolae» [18]. Using 82 points in the phase interval from -0.06 to +0.06, we got the value of $m_{min}=16.154\pm0.015^m$, corresponding to a «triangular» shape of the minimum. The accuracy estimate of the phase of the minimum is practically the same for trigonometric polynomial for $s=16$ (0.00119) and «asymptotic parabolae » (0.00122).

*Method of «slice approximations»*

Taking into account the locality of minima of a total duration $D$ [1], it is natural to split the phase curve into parts – the «out-of-eclipse» ($D/2, 0.5-D/2$; $0.5+D/2, 1-D/2$), «primary minimum» ($-D/2,+D/2$) and «secondary minimum» ($0.5-D/2, 0.5+D/2$). We assume thet the period and the initial epoch were determined using other methods (typically, for known stars, taking into account earlier observations of other authors), and the orbit of a binary star is circular (thus the primary and secondary minima have the same duration $D$, and are separated by a half of the period $P$). As usual, we suggest that the phases $\phi$ ($E+\phi=(t-T_0)/P$) from the interval $[0,1)$ may be extended to all real values by adding to them of the arbitrary integer numbers. The integer numbers $E$ are called the «cycle number » [e.g. 8,9].

Under such assumptions, we write

$x_C(\phi)=C_1+C_2\cos(2\pi\phi)+C_3\sin(2\pi\phi)+C_4\cos(4\pi\phi)+C_5\sin(4\pi\phi)+$

$+C_6H(\phi;C_8;\beta_1)+C_7H(\phi-0.5;C_8;\beta_2)$ (4)

Assuming a circular orbit, the width $D=2C_8$ of both minima is the same. However, generally, the depths $C_6$, $C_7$ and the profile parameters $\beta_1$ and $\beta_2$ are different for the primary and the secondary minima.

The usage of trigonometric polynomial of degree 2 for approximation of the "out-of-eclipse" part of the light curve is caused by physical models. The coefficients $C_2$, $C_4$ correspond to main contributions of the effects of «reflection» and «ellipsoidality», and $C_3$, $C_5$ describe the light curve asymmetry, which is usually related to the O'Connel effect (the spottedness of one or both stars) (e.g. [9,13]).

The main problem is to choose a dimensionless function $V(z)=H(\phi;C_8;\beta)$, which describes the profile of the minimum, where $z(\phi;C_8)= \phi /C_8$. For width-limited functions, for $|z|\geq 1$ we define $V(z)=0$, i.e. hereafter we deal with a determination of the function $V(z)$ only inside the interval [-1,1].

Obviously, due to a symmetry of the minimum, the function is to be symmetric $V(-z)=V(z)=V(|z|)$ and, optionally, $V(0)=1$. The requests for continuity and piecewise monotony of the light curve leads to conditions $V(\pm 1)=0$, $dV(|z|)/d|z| \leq 0$

In the monograph [12], there are listed classical «filter functions»: «rectangular» $V(z)=1$, «triangular» (Bartlett) $V(z)=1-|z|$, von Hann $V(z)=(1+\cos\pi z)/2$ et al. However, their profiles are significantly different from that observed in eclipsing stars. Besides, for description of the profiles of minima, it is optional to fasten computations and thus to simplify the function $V(z)$ and to minimize the number of additional parameters ($\beta$). The next class of functions used for the approximations, was

$$V(z) = (1-z^2)^\beta \qquad (5)$$

The family of functions $V(z)$ is shown in Fig.2 (left). For $\beta=0$, we get a classical «rectangular» shape, for $\beta=0.5$ – «semi-circle», for $\beta=1$ - parabola, for $\beta \to \infty$, a gaussian $V(z) =\exp(-\beta|z|^2)$.

An alternate class of functions was that proposed in 2010 by Andronov [19, http://www.astrokarpaty.net/kolos2010abstractbook.pdf]:

$$V(z) = (1-|z|^\beta)^{3/2} \qquad (6)$$

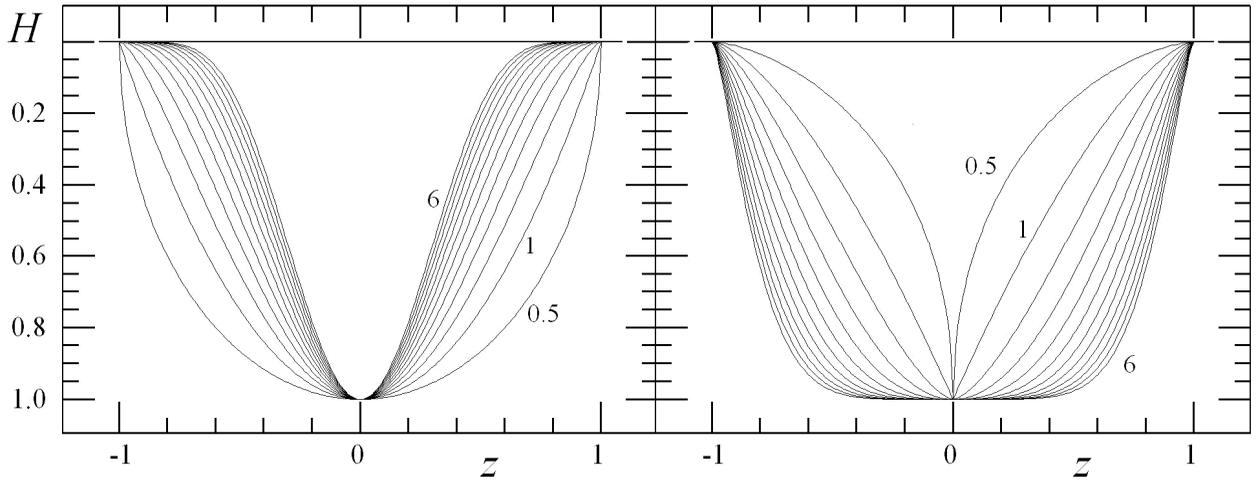

Fig. 2. Values of the function $H(z;\beta)$ for different $\beta$ from 0.5 to 6 with a step of 0.5 for Eq. (5) (left) and Eq.(6) (right).

The family of these functions is shown in Fig.2 (right). The fixed power index of 3/2 corresponds to an asymptotic shape close to the eclipse borders. For $\beta \to 0$, the function shows a decrease towards zero, what has no physical sense. For $\beta \to 1$, the shape is most close to the triangular one. With an increasing $\beta$, the shape becomes more flat in the middle, what corresponds to a full eclipse. Thus a preliminary qualitative analysis shows an applicability of the family of functions (6) for a description of phase light curves of the EA-type eclipsing binaries. As the method was developed for new Algol-type variables, at was called NAV (New Algol Variable).

*Determination of Parameters of Approximation*

In Eq. (4), the parameters $C_1 .. C_7$ may be determined using a linear least squares method, assuming fixed parameters $C_8 = D/2$, $C_9 = \beta_1$, $C_{10} = \beta_2$. With this, the «hidden unknown parameters» may be the initial epoch $T_0$ and the period $P$, which are used for the determination of the phases of observations.

The parameters $C_8$, $C_9$, $C_{10}$ themselves may be determined using any of the appropriate numerical methods – sequential search, Monte-Carlo, or differential corrections. For the last method, is is used to compute derivatives of the smoothing function with respect to these parameters. Intermediate derivatives from the shape function are equal to

$$dV/d|z| = -1.5\beta\, |z|^{\beta-1}\, (1-|z|^\beta)^{1/2}, \tag{7}$$

$$dV/d\beta = -1.5\, |z|^\beta\, (1-|z|^\beta)^{1/2}\, \ln |z|, \tag{8}$$

Although the simplicity of the used functions and corresponding derivatives, the program of computation of the smoothing function and its derivatives is rather cumbersome because of the piecewise approximation. The method of differential corrections (e.g. [10,11]) needs a determination of initial values of the parameters close to optimal values. Thus we realized a method of the sequential search of the parameters, and a further improvement of their values using the method of differential corrections.

For the phase curve of the object VSX0224, which was used for an example, we got the following values of the parameters: $C_8=0.1012$, $C_9=1.040$, $C_{10}=0.1$, $\sigma=0.0466^m$. According to the r.m.s. deviation of observations from the smoothing curve, such an approximation corresponds to a trigonometric polynomial of a degree with $s=10$ and the number of parameters $m=1+2s=21$.

The mean accuracy of the smoothing curve of $0.0065^m$ is by a factor of 1.5 better than that for the best (according to this criterion) trigonometric polynomial with $s=7$, and by a factor of 2 is better than for the optimal (according to the Fischer's criterion) approximation with $s=16$.

The depth of the primary minimum in respect to a second-order trigonometric polynomial $C_6=0.686\pm0.012^m$, and of the secondary minimum $C_7=-0.085\pm0.124^m$, what is zero within error estimates. The brightness at the primary minimum is $m_{min}=16.156\pm0.008^m$, at the maximum $m_{Maxn}=15.437\pm0.005^m$.

It should be noticed that the amplitude of the brightness variations $m_{minI}$-$m_{Max}$ does not coincide with the depth of the primary minimum $C_6$, because the latter parameter exhibits an brightness deficiency in respect to the «approximation of the "out-of-eclipse" part». This approach is close to the «rectification» of the light curve [9], but is used only for minima, and not for all the curve.

One more approximation of minimum was proposed in 2011г. By Mikulášek et al. [20], which, in our designations, may be written as:

$$V(z)=1-(1-\exp(-z^2))^\beta \tag{9}$$

For $\beta=1$, this function coincides with a Gaussian. However, as in the case of approximation (5), produces a family of functions. Obviously, fro this function the limitation $|z| \leq 1$ is absent, and formally the minimum has an infinite width. Thus the determination of the parameter $D$ for the «General Catalogue of Variable Stars» is impossible if using this function.

For VSX0224, the best approximation is at $C_8=0.0536$, $\sigma=0.0476^m$, $C_6=0.622\pm0.012^m$, what is significantly less than the value mentioned above; $m_{min}=16.093\pm0.007^m$, what is worse than the approximation using the trigonometric polynomial or the function NAV (4,6).

A complete form of the approximation [20] differs from ours. From a trigonometric polynomial, only a term with a cosine of a double phase remained (what corresponds to the effect of «ellipticity» only), but the O'Connell effect is described by a combination from 2 sine functions. Such an approximation describes the light curve even worse.

The smoothing light curves obtained using different methods are shown in Fig. 3. The best either for description of the "out-of-eclipse" part of the light curve, or of the eclipse, is the approximation using the NAV method (Fig. 3c).

Although for a "physical" modeling, the stellar magnitudes are usually transformed into relative intensity, and this signal can be approximated by the NAV method (including, for multicolor observations, and taking into account the precision of the individual observations), for listing the phenomenological characteristics in the catalogs, it is more suitable to use magnitudes.

*Light curve for a simplified model of spherical components.*

Not listing a complete set of formulae, which are used for computation of «physical» models of the light curves, which have been described in literature [9, 13-15] for many times, let us suggest the simplest case of spherical components (more described in the monograph [21]). We'll also neglect the limb darkening. For slightly elongated components, the method of «rectification» of the light curve was applied, where the intensity at a given phase was divided by a smoothed value, which was determined by approximation of the "out-of-eclipse" part of the light curve.

Let's denote the radii of two stars $R_1$ and $R_2$ ($\leq R_1$), the luminocities in a given filter $L_1$ and $L_2$ and the relative intensities $l_1=L_1/(L_1+L_2)$, $l_2=L_2/(L_1+L_2)$. Then the observed intensity $l(\Theta)=1$, if the visible distance between the centers of the stars $\Theta \geq R_1+R_2$ (out-of-eclipse part); $l(\Theta)=l_1$, if $0\leq \Theta \leq R_1-R_2$ (full eclipse, the second star is behind the first one); $l(\Theta)=1-l_1 (R_2/R_1)^2$, if $0\leq \Theta \leq R_1-R_2$ (partial eclipse, the first star is behind the second). In the interval $R_1-R_2\leq \Theta \leq R_1+R_2$, $l(\Theta)=1-l_1 S/\pi R_1^2$ (the first star is behind the second), $l(\Theta)=1-l_2 S/\pi R_2^2$ (the second star is behind the first one). Here $S=(\gamma_1-0.5\sin 2\gamma_1)R_1^2+(\gamma_2-0.5\sin 2\gamma_2)R_2^2$, where $\gamma_o=\arccos(\Theta^2+R_j^2-R_{3-j}^2)/(2\Theta R_j)$ – is the angle between the line of centers and the vectror from the center of the star number $j(=1,2)$ and the cross point of the borders of both stars in a projection onto the celestial sphere. The parameter $\Theta = a \sin\theta$ (where $\theta$ - is the angle between the line of centers and the line of view, $a$ – the distance between the centers of the stars) is related to the orbital inclination $i$ and phase $\phi$ with an equation $\sin^2\theta=1-\sin^2 i \cos^2(2\pi\phi)$ (e.g. [22]). Obviously, if $i=90^o$, the equation is simplified to $\theta=\arcsin|\sin(2\pi\phi)|$.

For small angles $\alpha$ we get an asymptotic relation $(\gamma-0.5\sin 2\gamma)\approx 2\gamma^3/3=2h^3/(3R^3)$, $S\approx 2h^3(1/R_1+1/R_2)/3$ and $\eta_1=(R_1+R_2)-\Theta \approx (R_1\gamma_1^2+R_2\gamma_2^2)/2=h^2(1/R_1+1/R_2)/2$, where $h=R_1\sin\gamma_1=R_2\sin\gamma_2\approx R_1\gamma_1\approx R_2\gamma_2$. Thus

$$S\approx 2^{5/2}\eta_1^{3/2} (1/R_1+1/R_2)^{-1/2}/3 \sim \eta_1^{3/2} \qquad (11).$$

As at the phase $\phi_e$ of the beginning (or, symmetrically, the end) of the eclipse, $\eta=0$, and the derivative differs from zero, then $\eta_1\approx (d\eta/d\phi)(\phi-\phi_e)$, and $S\sim(\phi-\phi_e)^{3/2}$. Namely this asymptotic relation was used when choosing the shape of the eclipse (6).

Before entering the phase of the full eclipse (second contact), we get a similar asymptotic relation

$$\pi R_2^2-S\approx 2^{5/2}\eta_2^{3/2} (-1/R_1+1/R_2)^{-1/2}/3 \sim \eta_2^{3/2}, \qquad (12)$$

where $\eta_2=\Theta-(R_1-R_2)$. Of course, this model can also be used in the program, and to determine 4 additional parameters $l_1(=1-l_2)$, $R_1/a$, $R_2/a$, $I$ instead of one $\beta$. The computational time becomes comparable with that needed for the "physical: models, thus such a «simplified model» is useless.

The derivative of the function (11) is equal to zero, if $\eta_1=0$, what means that the derivative of brightness on phase does not have a jump at moments of begin and end of eclipses. Thus, it increases very fast, thus it is possible to say on "borders of eclipses".

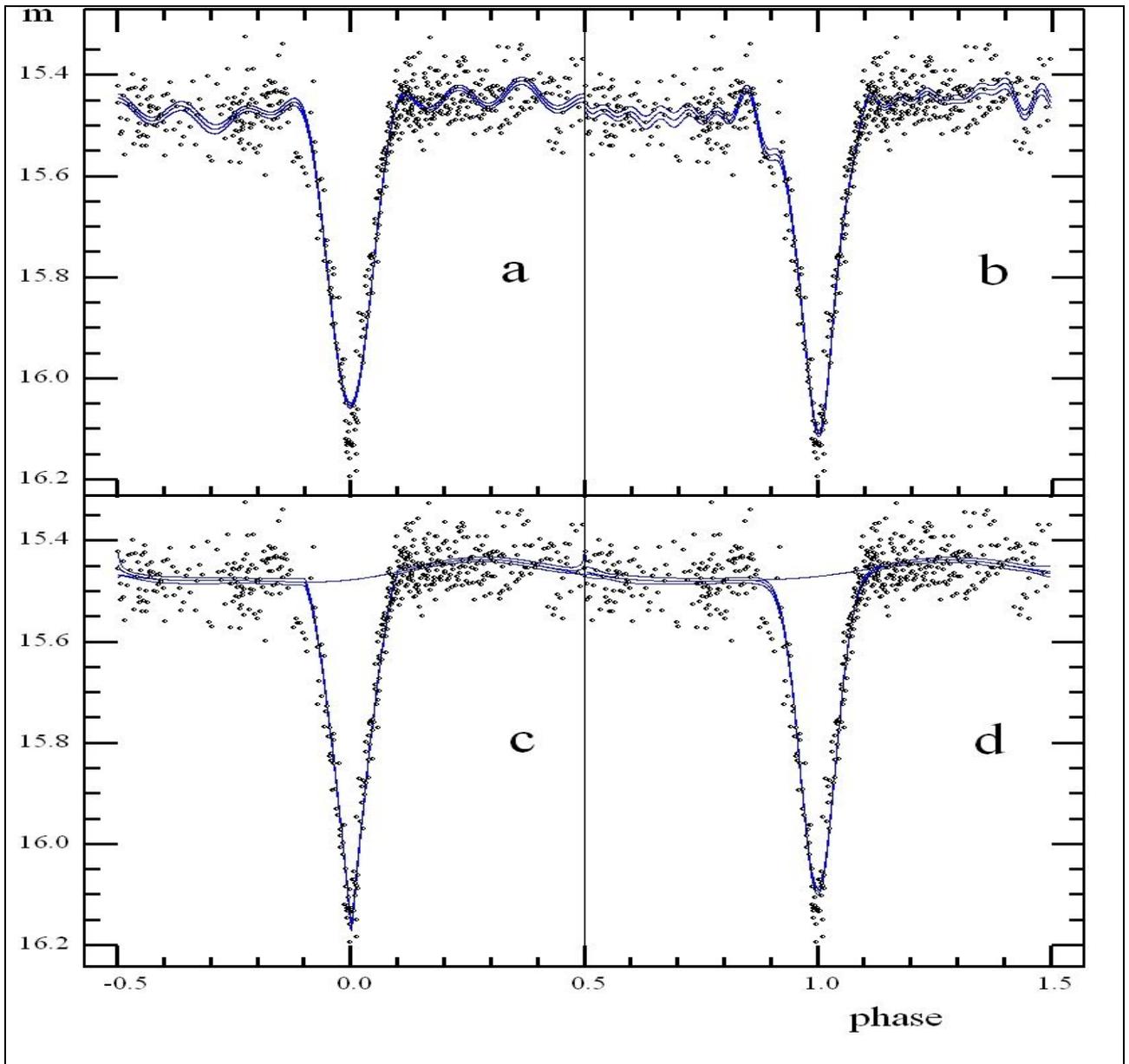

Fig. 3. Approximation of the phase curve of the eclipsing variable VSX0224: a) the trigonometric polynomial of degree $s=7$; b) $s=16$; c) using the method NAV; d) using the function (9) [20]. The error corridors for smoothing functions are shown. For the approximations (c,d) also shown are their extensions of the "out-of-eclipse part" onto the phases of eclipse.

It is possible to introduce an infinite number of monotonic functions, which follow the asymptotic dependencies for the beginning and end for a descending branch of an eclipse. Such properties are characteristic e.g. for a function

$$V(\phi)=\kappa w(\chi)(1-\chi)^{3/2}+(1-\kappa)(1-(1-w(\chi))\chi^{3/2}) \qquad (13)$$

where $w(\tilde{\chi})$ is a weight function, equal to 1 at $\chi=0$ and 0 at $\chi=1$, and its first and second derivations are equal to zero at the borders of the interval [0,1]. Here $\chi=(|\phi|-\phi_2)/(\phi_1-\phi_2)$, where $\phi_1$ and $\phi_2$ – are phases of the end of the eclipse and of the end of the total eclipse, which are also are unknown parameters, as well as $\kappa$. The parameter $\kappa$ is related to a difference in the proportionality coefficients in Eq. (11) and (12). Assuming a symmetry of the function $w(\chi)$ in respect to the mid-interval ($\chi=0.5$), one may suggest e.g. a simplest polynomial function with these properties:

$$w(\chi) = 1-10\chi^3+15\chi^4-6\chi^5. \qquad (14)$$

The function (13) well describes an ascending (and a symmetric descending) branch. However, according to a number of parameters, it is not suitable for the approximation. Thus instead of this model, we used the shape of the minimum (6) with only one phenomenological parameter $\beta$.

### Hypothesis of a Double Period for VSX0224

Despite the electronic publications on discovery of variability of VSX0224 contain information on the orbital period of $P=0.5227^d$, no secondary minimum near phase 0.5 within the error estimates is surprising, because it is an argument in favor of the fact that the surface brightness of the secondary component is negligible compared to that for the primary component, i.e. the secondary has a much lower temperature. The model of the system could be checked in future when carrying out spectral and multi-color observations.

If having one-color observations only, as in a case of VSX0224, one could make an additional test. Let's assume that the true period is twice larger: $P=2 \cdot 0.522724^d=1.045448^d$, and the primary and secondary minima are comparable in depth. Then, using the NAV procedure, we get a new approximation with a slightly smaller scatter $\sigma=0.0457^m$. According to the Fischer's criterion, the probability of random decrease to such a value is 33%, i.e. twice larger is a probability of an alternate hypothesis of non-random decrease. The corresponding set of the parameters: $C_8=0.0528$ (slightly more than a half of the value of $C_8=0.1040$ for the mentioned above model of "short period"), $C_9=1.030$ (in an agreement to the result mentioned above), $C_{10}=0.940$ (naturally, with a well defined secondary minimum, the value had become closer to $C_9$). The brightness at the minima $16.158\pm0.008^m$ (phase 0.0), $16.203\pm0.016^m$ (phase 0.5). One should note a brightness decrease at the "out-of-eclipse" part of the light curve near phase 0.86, what may be interpreted by a cold spot at one of the components of the binary system.

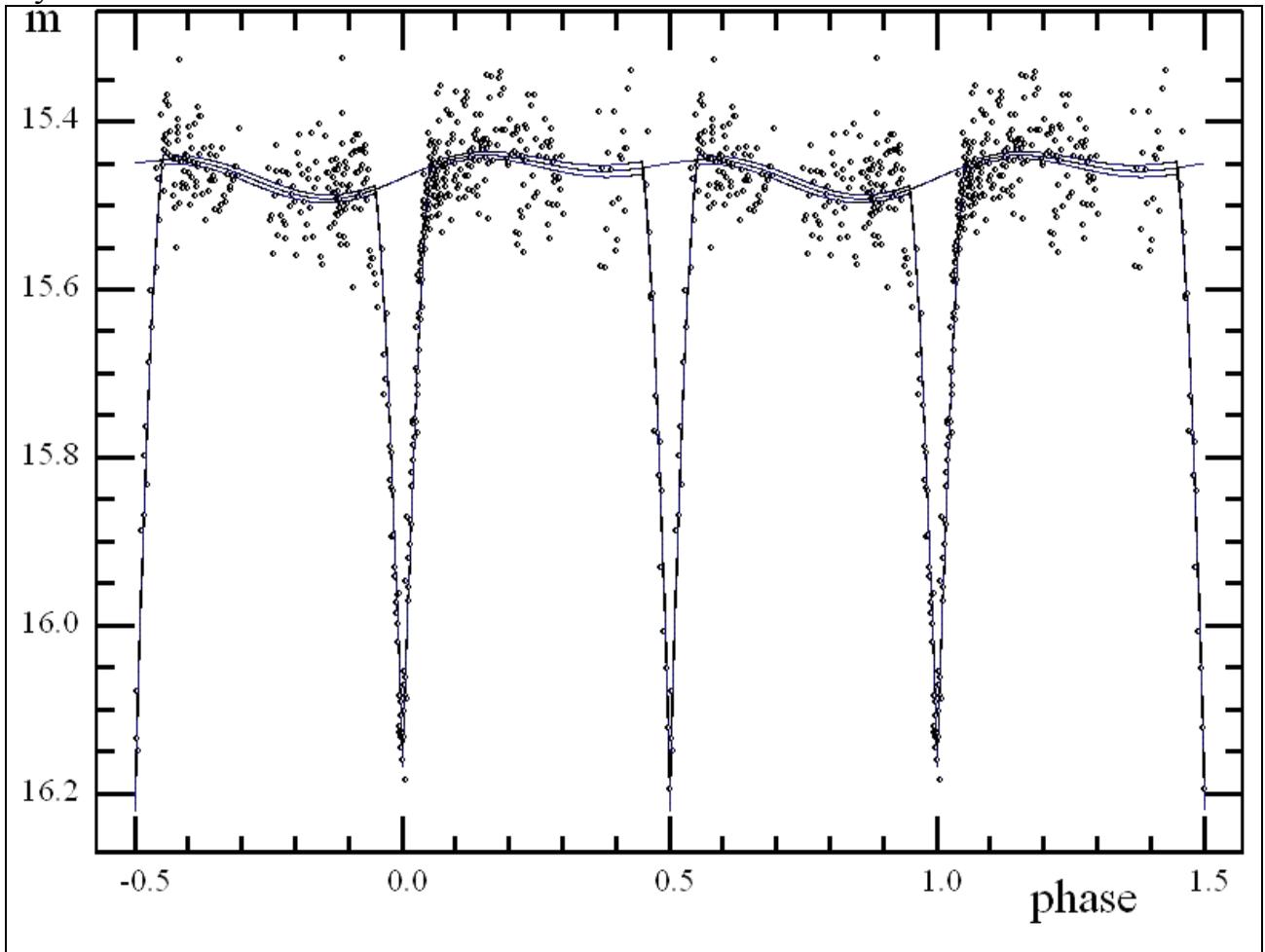

Fig.4. The phase light curve of VSX0224 for the hypothesis of «two minima close in depth» (or «double period») with the initial epoch $T_0$= HJD 2455106.3236 and period $P=2\cdot0.522724^d=1.045448^d$ and its approximation using the NAV method.

The depths of the minima $C_6=0.692\pm0.010^m$ and $C_7=0.745\pm0.019^m$, i.e. more deep (main) minimum in this model is the one at the phase 0.5, thus for listing in the catalogue one should use the initial epoch $T_0=2455106.8463\pm0.0006^d$. The depth difference $C_7-C_6=0.062\pm0.021^m$ is $3\sigma$, and may be declared as statistically significant.

Taking into account the Pogson's formula, at the minimum is eclipsed a part of the flux

$$l = 1 - 10^{-0.4C} \qquad (10)$$

where $C=C_6$ for the minimum at phase 0.0 and $C=C_7$ for the minimum at phase 0.5. For the mentioned above values, $l_1=0.4713\pm0.049$, $l_2=0.4965\pm0.0089$. The value of $l_1+l_2=0.9678\pm0.0101$ is very close to unity, arguing that both eclipses are «almost full». The ratio of the mean surface brightnesses $l_2/l_1=0.949\pm0.020$ is also close to unity within error estimates. An example of similar analysis for another system WZ Crv we presented in [23].

Thus we make a simplest suggestion that the components of the eclipsing system are practically the same, and an inclination of the orbit is close to 90°. In this case, one may use a relation $(R_1+R_2)/a=\sin\pi D=\sin 2\pi C_8$ [9,23], where $R_1, R_2$ – are radii of the components, and $a$ – the distance between their centers.

Also we suggest that both components are main sequence stars. For near-solar masses, $R/R_{Sun}=M/M_{Sun}$, so for the parameter $C_8=0.0528$ we get $M/M_{Sun}=0.835$, corresponding to the spectral class G8V [24]. Obviously, this is a preliminary estimate, which was obtained using a series of simplifying suggestions.

By interpolating the dependence «spectrum-mass-radius» from table 15.8 [24], we get (assuming «same stars»), $M/M_{Sun}=0.735$, $R/R_{Sun}=0.779$, Sp K2.8 V. Such a discrepancy of the corresponding parameters is related to differences of empirical «mass-radius» relations using different literature data.

Assuming that the larger star is also more hot (in an assumption of a main sequence), one may suggest that the eclipse at the phase 0.0 is a full one, and that at the phase 0.5 – a partial one. However, for a "serious" checking of this assumption, new more accurate observations are needed. Besides, a slight difference of surface brightness, arguing for difference in temperatures, should cause variations of the color index in the intervals of phases of eclipses with a maximum near phase 0.0 and a minimum near phase 0.5.

Let's estimate the contribution of «correction», taking into account the observed value of the ratio of the mean brightness, using linear interpolation in the interval of spectral classes K0-K5 of the masses, radii, absolute stellar magnitudes in the photometric system R, and of the parameter inverse to temperature ($1/T$). The resulting pairs of parameters for the observed values of the period and eclipse width: $M_1=0.740 M_{Sun}$, $M_2=0.732 M_{Sun}$, $R_1=0.785 R_{Sun}$, $R_2=0.775 R_{Sun}$, $T_1=4751$ K, $T_2=4696$ K, Sp1 K2.5V, $L_1/(L_1+L_2)=0.519$ (what is close to the observed value of 0.529). The absolute stellar magnitude $M_{R1}=5.81^m$ corresponds to a distance of ~ 207 pc. Once nore, as the difference between the minima depths is at ~ $3\sigma$, at the limit of detection, these estimates are listed for an illustration of difference in the parameters comparatively to the case of «two equal stars».

*Application to Other Binary Systems*

The method NAV was illustrated above by an application to one binary system VSX0224. However, it was used also for the stars with more wide minima – of the types EB, EW. In Fig. 5 is shown an approximation using the NAV method of the phase light curve BM UMa using the CCD V observations of Virnina et al. [25]. The system was classified as EW [26], and the «physical» modeling [25] using the program [27] confirms that this is an over-contact system. According to a «classical» definition [1], in such systems "it is impossible to specify the exact times of onset and end of eclipses". However, at the light curve is clearly seen a significant difference between the «smooth» variations, caused by ellipticity of components, and a relatively narrow minima. Statistically optimal are the parameters $C_8=0.104$, $C_9=1.4$, $C_{10}=1.1$.

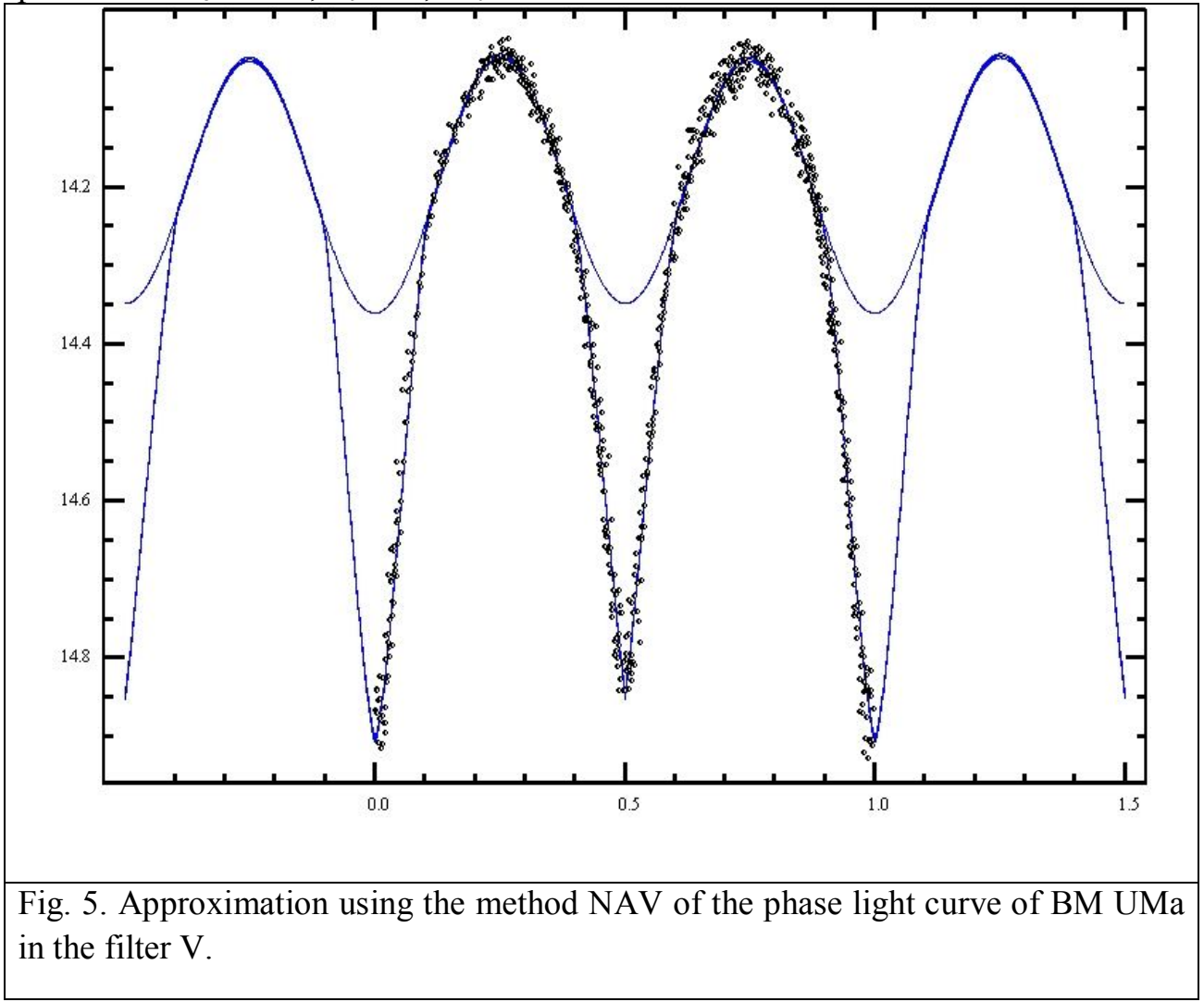

Fig. 5. Approximation using the method NAV of the phase light curve of BM UMa in the filter V.

Thus the method NAV is applicable also to the EW-type stars. But, in many cases their light curves are well approximated using a trigonometric polynomial [10,13,17]. Particularly, the optimal value for BM UMa is $s=8$, which was used In the paper [25]. ]. However, the number of parameters $m=1+2s=17$ is siglificantly larger than 10 in the NAV method, thus the statistical accuracy of the NAV approximation is by a factor of 1.3 better than for the trigonometric polynomial.

As a «precessor» of the method NAV, we have used a separate approximation of the "out-of-eclipse" and "complete" phase light curves of WZ Crv [23] using trigonometric polynomials of $s=2$. However, the method NAV gives a better approximation of the complete light curve with a determination of the borders of eclipse using the mathematical methods instead of subjective visual ones.

It should be noted a good agreement of the NAV approximation with theoretical curves computed using the Wilson-Devinney [14] method. We have used the program elaborated by S.Zoła [27].

**Discussion**

For the determination of the parameters of the light curves ov variable stars, which are needed for the registration in the «General Catalogue of Variable Stars», it is needed to compute statistically optimal approximations, thus we have introduced the method "NAV", which was applied to observations of dozens of real stars and their theoretical models. Such a phenomenological modeling is especially effective for new (newly discovered or poorly studied) variables of the Algol type (New Algol Variable), which, particularly, are studied in a frame of projects «Ukrainian Virtual Observatory» [28] and «Inter-Longitude Astronomy » [29].

With a decreasing width of the eclipse, the number of parameters of harmonics increases and may reach few dozens. The Gibbs phenomenon and statistical errors of measurements lead to occurrence of short-period apparent waves, which physically are not present in the signal. We propose the method of decrease of statistically optimal number of parameters consisting in an alternate choice of a group of the basic functions, which are different from a classical set of sines and cosines. We use the combination of the second-order trigonometric polynomial (TP2, what describes the effects of "reflection", "ellipsoidality" and "spottedness"), and of localized contributions of minima (parametrized in depth and shape separately for the primary and the secondary minimum). This allows to significantly improve the quality of approximation.

The main parameters to describe eclipses are the half-width $C_8=D/2$, which is the same for primary and secondary minima in an often assumption of circular orbit, as well as the parameters $\beta_1$ and $\beta_2$, which describe the eclipse shapes. The numerical models show that, for the stars close in dimensions, $\beta_1 \approx \beta_2 \approx 1$, and the shape at the center of eclipse is a «triangular». For the case of transition of a smaller star in front of the disc of a larger one («transit»), the main effect of brightness variations is due to the limb darkening of the eclipsed star, and one may expect that $\beta_1 \approx 2$. In this case, another eclipse is full, and $\beta_2 > 2$.

A special interest attract the full eclipses. Formally they are described by the shape (6) with relatively large values of the parameter β. For a full eclipse, one may suggest more complicated models mentioned above. However, they suggest a larger number of parameters and resources for computations, having practically the same quality of approximation. Thus we have not used them.

The eclipse shape describes with function (6) produces better agreement with the «physical» models than (5) or (9).

The method "NAV" is applicable also to the types EB and EW, therefore being most effective for the EA type.

Using results of analysis, we estimated characteristics of VSX J022427.8-104034.

*Acknowledgments:* The author thanks to N.N.Samus', V.Yu.Terebizh and V.P.Grinin, D.Chochol, S.Otero for useful discussions, S.Zoła for the program for physical modeling of eclipsing binary stars, described in [27], and the Queen Jadwiga Foundation, Jagiellonian University (Krakow) for an individual grant. The work was initiated by numerous discoveries of new eclipsing binary stars by N.A.Virnina, a graduate student, who published 19 articles under my supervision.